\documentclass{appolb}
\usepackage{amsmath}
\usepackage{amssymb} 
\usepackage[small]{caption2} 
\usepackage{fleqn} 
\usepackage{graphicx} 
\usepackage[small,loose]{subfigure}  
\usepackage{cite} 
\advance \headheight by 3.0truept       
\DeclareMathOperator{\re}{Re}

\newcommand{\CenterObject}[1]{\ensuremath{\vcenter{\hbox{#1}}}}
\newcommand{\D}{\mathrm{d}}

\newcommand{\SU}[1]{\ensuremath{\mathrm{SU}(#1)}}

\begin{document}
\preprint{}
\title{\appHuge{A Note on  Fine--Tuning\\
in Mirage Mediation}%
\thanks{Contributed to the proceedings of GUSTAVOFEST: 
Symposium in Honor of Gustavo C.\ Branco: 
CP Violation and the Flavour Puzzle, Lisbon, Portugal, 19-20 Jul 2005.}%
}
\author{Oleg Lebedev,
Hans Peter Nilles,
Michael Ratz
\address{Physikalisches Institut der Universit\"at Bonn, Nussallee 12, 
53115 Bonn, Germany}
}
\maketitle
\begin{abstract}
Recent progress in string theory moduli stabilization   has motivated   a  mixed
modulus--anomaly mediated supersymmetry breaking scenario, also dubbed `mirage
mediation'. This scenario has a number of phenomenologically  attractive
features, in particular with respect to  the cosmological gravitino/moduli
problem. In this note, we investigate the issues of fine--tuning  associated
with obtaining the correct electroweak symmetry  breaking scale in the mirage
mediation scenario.  We find that, due to lighter gluinos, the fine--tuning is 
smaller than that in other mediation mechanisms.  
\end{abstract}
  
\section{Introduction}

The minimal supersymmetric extension of the standard model (SM), the MSSM,
enjoys high popularity. One of the main theoretical reasons for it is that a
supersymmetry (SUSY) breaking scale close to the electroweak (EW) scale would
allow one  to understand  stability of the electroweak scale against radiative
corrections. In addition, SUSY models with TeV--scale soft masses offer the most
attractive scenario for  perturbative gauge coupling unification.

On the other hand, it is precisely the Higgs sector  that casts some shadow  on
this scheme.  The reason is the following. At tree-level, one has an upper bound
on the mass of the lightest Higgs \cite{Inoue:1982ej,Flores:1982pr},
\begin{equation}
 m_{h^0}~<~m_Z\,|\cos2\beta|\;,
\end{equation}
where, as usual, $\tan\beta=v_u/v_d$ denotes the ratio of the two Higgs
expectation values. This is in conflict with the current experimental lower
limit on the Higgs mass,
\begin{equation}\label{eq:LowerBoundHiggsMass}
 m_\mathrm{Higgs}~\gtrsim~114\,\mathrm{GeV}\;.
\end{equation}
Luckily, this does not rule out the MSSM because there are sizable radiative
corrections to the Higgs mass \cite{Haber:1990aw,Okada:1990vk,Okada:1990gg}, the
most important one being
\begin{equation}
 \Delta(m_{h^0}^2)
 ~\simeq~
 \frac{3 g^2 m_t^4}{8\pi^2 m_W^2}
 	\,\ln\left(\frac{m_{\tilde{t}_1}\,m_{\tilde{t}_2}}{m_t^2}\right)
 \;,
\end{equation}
where  $m_{\tilde{t}_{1,2}}$ denote the masses of the scalar top quarks
(`stops').  To lift
the Higgs mass above the lower experimental bound
\eqref{eq:LowerBoundHiggsMass}, one needs sizable superpartner masses,
$m_{\tilde{t}_{1,2}}\gtrsim\mathrm{TeV}$. 

On the other hand, electroweak symmetry breaking requires
\begin{equation}\label{eq:mZ2}
 \frac{m_Z^2}{2}
 ~=~
 -\mu^2+\frac{m_{H_d}^2-m_{H_u}^2\,\tan^2\beta}{\tan^2\beta-1}\;,
\end{equation}
where $m_{H_{d,u}}$ are the soft Higgs mass parameters and $\mu$ is the
supersymmetric $\mu$--parameter, both evaluated at the electroweak scale.
Thus one naturally expects the SUSY mass parameters (or at least $m_{H_u}$
and $\mu$) to be of order $100\,\mathrm{GeV}$, otherwise a large cancellation 
between {\it a priori} independent terms would be required. 
The problem is, however, that the GUT scale  parameters get renormalized, in
particular
\begin{equation}\label{eq:RGEmHu2a}
 \frac{\D}{\D t}m_{H_u}^2
 ~\supset~
 \frac{3\,y_t^2}{8\pi^2}\,
 (m_{\widetilde{t}_\mathrm{R}}^2+m_{\widetilde{q}_\mathrm{L}^{(3)}}^2)
 \;,
\end{equation}
where $t$ is a scale variable, and   $m_{\widetilde{t}_\mathrm{R}}^2$ and
$m_{\widetilde{q}_\mathrm{L}^{(3)}}^2$ denote the soft mass parameters for the
right- and left-handed $\widetilde{t}$, respectively. Thus, one has
a radiative correction
\begin{equation}\label{eq:deltaRGmHu2}
 \delta_\mathrm{_{RG}}\,m_{H_u}^2
 ~\simeq~
 -\frac{3\,y_t^2}{8\pi^2}\,
 (m_{\widetilde{t}_1}^2+m_{\widetilde{t}_2}^2)\,
 \ln\left(\frac{\Lambda}{m_{\widetilde{t}}}\right)\;,
\end{equation}
with $m_{\widetilde{t}}=\sqrt{m_{\widetilde{t}_1}\,m_{\widetilde{t}_2}}$ and
$\Lambda$ being the high energy scale at which the boundary conditions are specified.
Taking $\Lambda= M_\mathrm{GUT}\simeq2\times10^{16}\,\mathrm{GeV}$, one  expects
$|\delta m_{H_u}^2|$ to be of  order $m_{\widetilde{t}}^2$. The former is
constrained by the Higgs mass bound to be in the $(\mathrm{TeV})^2$ range,
indicating that large cancellations in Eq.~\eqref{eq:mZ2} are required. This is
the essence of the supersymmetric `little hierarchy' problem
\cite{Chankowski:1997zh,Barbieri:1998uv}.

There are  two obvious ways to evade this conclusion:
\begin{enumerate}
 \item  low $\Lambda$ 
 \item  cancellations 
\end{enumerate}
The first possibility has been studied rather extensively (see, e.g.,
\cite{Casas:2003jx}). Although the fine--tuning can be superficially reduced,
there is a price one has to pay. Namely, one loses the apparent gauge coupling
unification and other appealing features involving high scales such as the
see-saw mechanism (see, e.g., \cite{Casas:2004gh}). 

The second possibility gained some popularity more recently. It has been
realized that in the framework of flux compactifications of string theory,  the
soft masses \cite{Choi:2004sx,Choi:2005ge} have certain features that may help
ameliorate the SUSY fine--tuning problem. In particular, in the scheme of 
`mirage mediation'\footnote{This name was coined in
\cite{Loaiza-Brito:2005fa}.}, some cancellations between the input value of
$m_{H_u}^2$ and its RG corrections are  possible
\cite{Choi:2005hd,Kitano:2005wc}. In what follows, we focus on this possibility.

\section{Mirage Mediation and Mirage Unification}
\label{sec:MirageMediation}

We start by reviewing main features of the `mirage mediation' scheme. It is
motivated by Calabi--Yau compactifications of string theory with fluxes. A
particular realization is given by the model of  Kachru, Kallosh, Linde and
Trivedi (KKLT) \cite{Kachru:2003aw}. In this scenario, all the moduli are fixed
and    the cosmological constant is close to zero. One of the  
phenomenologically attractive features of this setup is  a  hierarchy among the
MSSM soft masses, the gravitino  and  moduli masses
\cite{Choi:2004sx,Choi:2005ge},
\begin{equation}
m_{\rm _{MSSM}} \ll m_{3/2} \ll m_{\rm moduli} \;,
\end{equation} 
such that the gravitino and   the moduli can be made heavy so as to avoid
cosmological problems  associated with late decays of these particles.

Another interesting  feature is that the MSSM  soft terms receive comparable
contributions from  gravity (modulus) mediated  and  anomaly mediated  
\cite{Randall:1998uk,Giudice:1998xp} SUSY breaking. Specifically, for  the MSSM
on D7 branes we have \cite{Choi:2005ge}\footnote{We follow the conventions of
\cite{Falkowski:2005ck}. Here  we choose the `effective modular weights' $n_i=0$
for the matter fields, whereas other choices are also possible (see e.g.
\cite{Choi:2005hd,Kitano:2005wc}). We also assume that the Yukawa couplings are
independent of the $T$--modulus. } 
\begin{eqnarray} 
\label{e.mmast} 
 M_a &=& M_s\, \left [\alpha +  b_a\, g_a^2 \right ] \;, 
\nonumber \\
m_i^2 &=&  M_s^2\, \left [ 
\alpha^2  - \dot \gamma_i  + 2 \alpha\, (T + \bar T)\, \partial_T  \gamma_i   
\right ] \;, 
\nonumber \\
A_{i j k} &=& M_s\, \left [  3 \alpha - \gamma_i - \gamma_j -\gamma_k \right ]  \;. 
\end{eqnarray} 
Here $M_s=m_{3/2}/(16 \pi^2)$ is the scale of the soft terms, $\alpha$ measures
the balance between the anomaly and the $T$-modulus mediated contributions and 
typically lies in the  range $0< \alpha\le10$, $b_a$ are the beta function
coefficients for the gauge couplings $g_a$, $\gamma_i$ is the anomalous
dimension and  $\dot{\gamma}_i = 8 \pi^2 {\partial  \gamma_i \over \partial \log
\mu}$. The modulus $T$ is responsible for the standard model gauge couplings
and  is fixed at the value $\re T=1/g^2_\mathrm{GUT} \simeq 2$. 

It has been observed that the above  gaugino masses unify at an intermediate, the so-called
`mirage unification' scale \cite{Choi:2005uz,Endo:2005uy},
\begin{equation}
\mu_{\rm mir}  = M_{_{\rm GUT}} {\rm e}^{-8\pi^2 / \alpha} \;.  
\end{equation} 
This is a feature of the boundary conditions in the 
mixed anomaly--modulus mediated scenario. At the mirage scale no `new' physics appears,
hence the name `mirage'. We note that large $\alpha$ correspond to modulus 
domination and the mirage scale is the GUT scale, whereas in  the anomaly dominated
limit  $\alpha \rightarrow 0$ the mirage scale approaches zero. 

\begin{figure}[!t]
\centerline{
\subfigure[$M_i(\mu)$.\label{fig:MirageUnificationMi}]{\CenterObject{\includegraphics[scale=0.75]{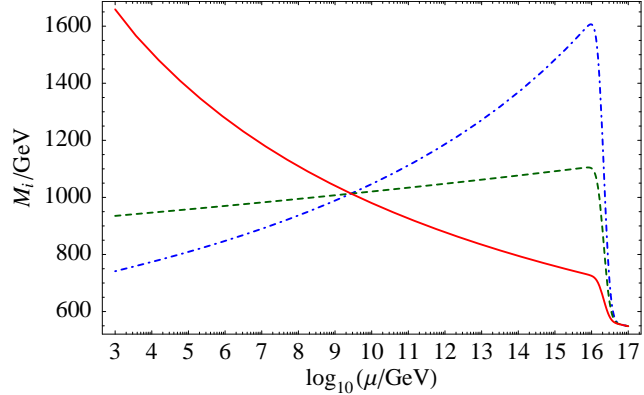}}}
}
\centerline{
\subfigure[$\alpha_i(\mu)$.]{\CenterObject{\includegraphics[scale=0.75]{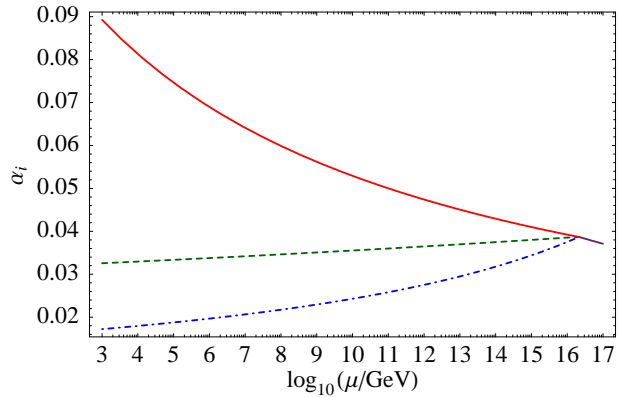}}}
 }
 \caption{The `mirage' and `real'  unification of the gaugino masses (a),
 and the gauge coupling unification (b). 
 The solid (red), dashed (green), dash-dotted (blue) curves in Plot (a) show the evolution of
 $M_3$, $M_2$, $M_1$. 
 Plot (b) displays  the evolution of  $\alpha_i=g_i^2/4\pi$.
 Above the unification scale $M_\mathrm{GUT}\simeq2\times10^{16}\,\mathrm{GeV}$,
 we use an \SU5 theory with 3 generations of
 $\boldsymbol{10}\oplus\overline{\boldsymbol{5}}$, one pair
 $\boldsymbol{5}+\overline{\boldsymbol{5}}$ as well as an adjoint \SU5
 Higgs.}
 \label{fig:MirageUnification}
\end{figure}

One might now ask the question what happens if we start with a really unified theory
such as an \SU5 GUT.
Clearly, at energies above the unification scale, one has
only one gaugino and  there is only one gaugino mass. 
Just below the GUT scale we have non--universal gaugino masses due to the anomaly 
contribution. These two limits are reconciled via the  threshold corrections at the
GUT scale (cf.\ e.g.\ \cite{Luty:2005sn}).
We are therefore led  to the  picture where there is  `real
unification' of gaugino masses in the realm of the unified theory, and  `mirage
unification' at an intermediate scale. We illustrate these effects in
Fig.~\ref{fig:MirageUnification}.

An important feature of the setup is that for typical $\alpha$ the gluino is the
lightest gaugino at the GUT scale. This is because the $\SU{3}_c$ beta function
is the largest one and it is being subtracted form the universal modulus
contribution.  Clearly, this feature is specific to `mirage mediation' and
does not hold in its  limiting cases $\alpha \rightarrow 0$, $\alpha \rightarrow
\infty$.  Light gluinos are desirable regarding  the fine--tuning problem since 
gluinos control to a large extent the RG running of the squark and Higgs masses.
In the next section, we analyze this issue in detail.

\section{Fine-Tuning in Mirage Mediation}

As discussed in the Introduction, a certain degree of fine--tuning is required
in SUSY models to obtain the electroweak scale from the scale of the soft
masses. This  can be understood qualitatively  by using approximate analytic
solutions to the RG equations.  One can rewrite Eq.~\eqref{eq:mZ2} in terms of
the input SUSY parameters {\it at the GUT scale} as \cite{Kane:2002ap} 
\begin{eqnarray}
 m_Z^2 
 &\simeq& -1.8\,\mu^2 +5.9\, M_3^2 -0.4\, M_2^2 -1.2\, m^2_{H_u} 
                   +0.9\, m^2_{q_\mathrm{L}^{(3)}} 
				   +0.7\, m^2_{u_\mathrm{R}^{(3)}}  \nonumber\\
 &&{}-0.6\, A_t\, M_3 + 0.4\, M_2\, M_3+ \dots\;,   \label{eq:Kane}
\end{eqnarray}
where we have taken $\tan\beta=5$ and neglected terms with smaller numerical
coefficients. This equation shows sensitivity of $m_Z$ to various input
parameters. If all the parameters are about $100\,\mathrm{GeV}$, no significant
fine--tuning is needed. However, as we have argued the lightest Higgs  mass
bound requires the stops of  about $1\,\mathrm{TeV}$ (at the EW scale), such
that $100\,\mathrm{GeV}$ input parameters are typically inconsistent with
experiment. Then, in order to get the right $m_Z$, some cancellations in
Eq.~\eqref{eq:Kane} are needed.

\subsection{Cancellations and Tachyons}
  
Clearly, $m_Z$ is most sensitive to the input value of the gluino mass. Thus 
reduction of $M_3$ is welcome from the fine--tuning perspective, as it occurs in
`mirage mediation'. Then given a larger  $m^2_{H_u}$ at the GUT scale,
significant cancellations in Eq.~\eqref{eq:Kane} are possible to achieve. In
other words, $m^2_{H_u}$($m_Z$) in Eq.~\eqref{eq:mZ2} can be  made of order
$100\,\mathrm{GeV}$ by cancelling its GUT input value by the RG evolution. 
However, implementation of this mechanism  in simple versions of `mirage mediation'
requires tachyons at the GUT scale and is strongly constrained
by the Higgs mass bound.

\begin{figure}[!t] 
 \subfigure[$m^2_{\widetilde{\ell}^{(i)}_\mathrm{L}}$.]{%
 	\CenterObject{\includegraphics[scale=0.9]{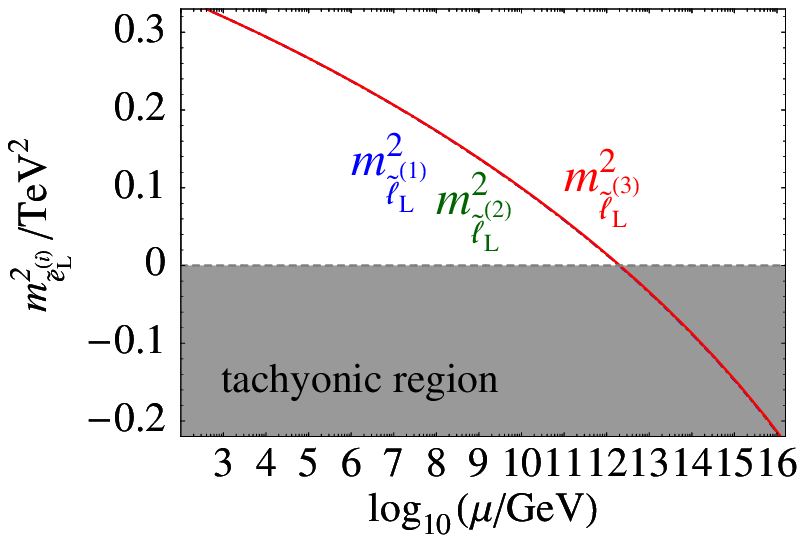}}}\hfil
 \subfigure[$m^2_{\widetilde{e}^{(i)}_\mathrm{R}}$.]{%
 	\CenterObject{\includegraphics[scale=0.9]{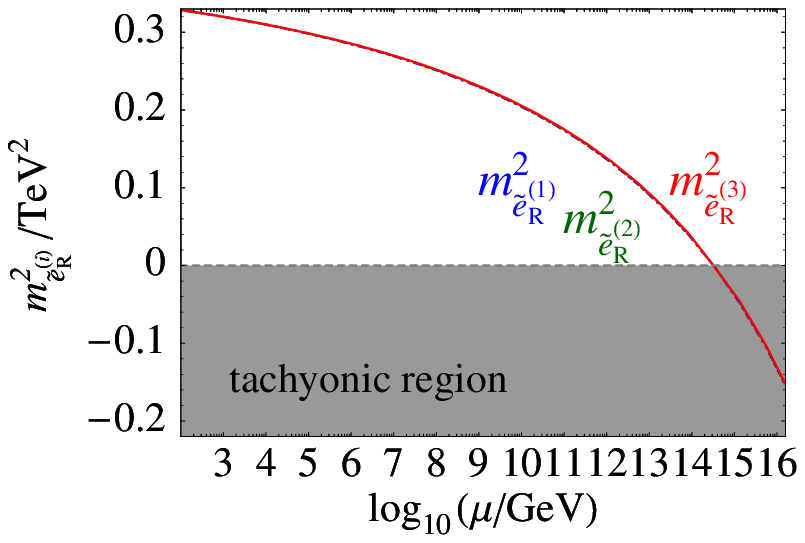}}}\\
 \subfigure[$m^2_{\widetilde{q}^{(i)}_\mathrm{L}}$.]{%
 	\CenterObject{\includegraphics[scale=0.9]{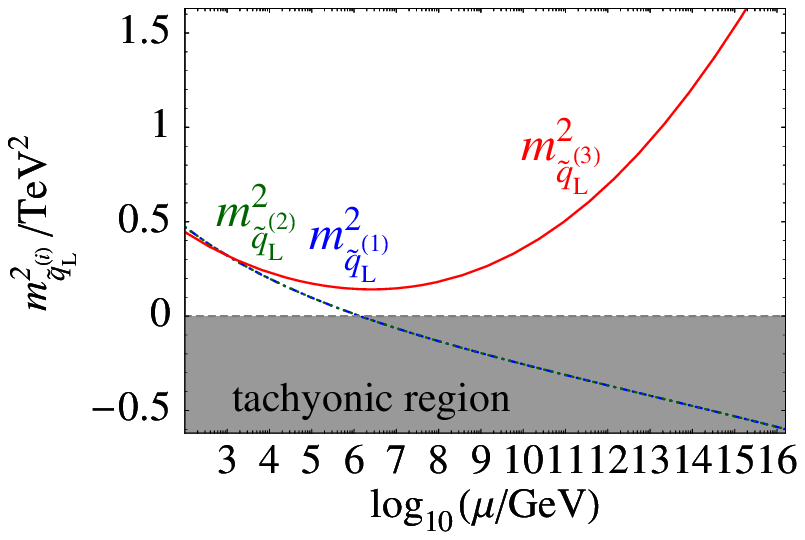}}}\hfil 
 \subfigure[$m^2_{\widetilde{u}^{(i)}_\mathrm{R}}$.]{%
 	\CenterObject{\includegraphics[scale=0.9]{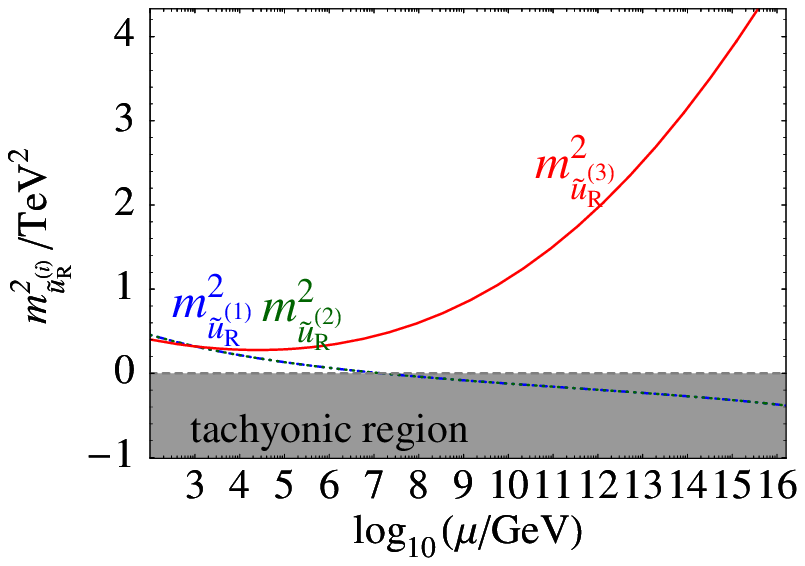}}}\\
 \subfigure[$m^2_{\widetilde{d}^{(i)}_\mathrm{R}}$.]{%
 	\CenterObject{\includegraphics[scale=0.9]{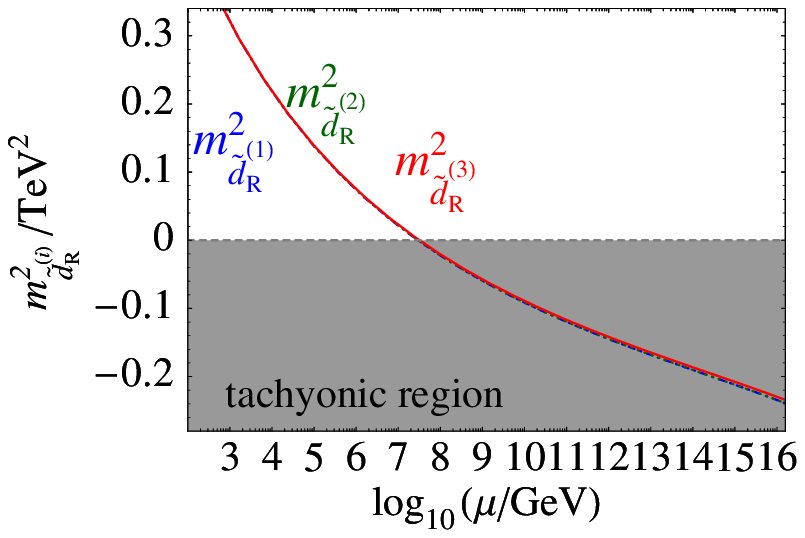}}}\hfil 
 \subfigure[$m_{H_i}^2$.]{%
 	\CenterObject{\includegraphics[scale=0.9]{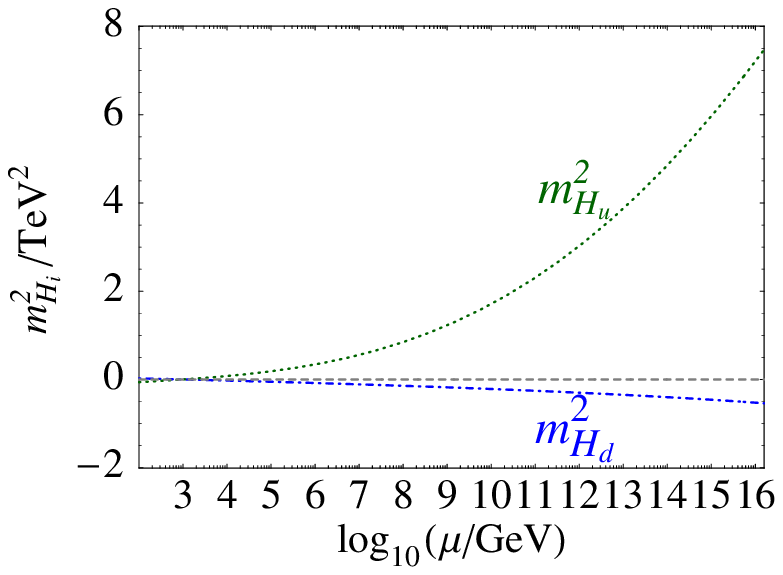}}}\hfil
\caption{RG evolution of the soft masses for $\mu_\mathrm{mir} \simeq 1\,\mathrm{TeV}$
and $\tan\beta=5$.
The GUT
boundary conditions are chosen such that $M_i=A_i=\sqrt{2} m_{\rm sferm.} \sim
\mathrm{TeV}$ and $m_{H_{u,d}} \ll \mathrm{TeV}$ at the TeV scale. }	
\label{fig:plots}
\end{figure}

An example of the cancellation effect is shown in Fig.~\ref{fig:plots}. There
the GUT boundary conditions are chosen such that $M_i=A_i=\sqrt{2} m_{\rm
sferm.} \sim \mathrm{TeV}$ and $m_{H_{u,d}} \ll \mathrm{TeV}$ at the TeV scale
\cite{Choi:2005hd,Kitano:2005wc}.  Clearly, this requires tachyonic squarks and
sleptons at the GUT scale signalling that the boundary conditions are not well
defined. One may ignore this problem by saying that what matters is the masses
at the `mirage scale', but as we have discussed there is no new physics
appearing at this scale and the `true' boundary conditions should be defined
at the GUT scale, above which a new theory sets in.  The presence of tachyons
appears as too high a price for ameliorating fine--tuning.

On the other hand, in general, large cancellations in Eq.~\eqref{eq:Kane} can be
achieved without tachyons in the spectrum. What matters for the fine--tuning
problem is the stop masses and these are non--tachyonic  (Fig.~\ref{fig:plots}).
The other mass squareds can be made positive by choosing appropriate `effective
modular weights' for them, subject to the FCNC constraints. This will lead to a
more complicated set of boundary conditions reminiscent of the  general MSSM.
Given  enough freedom one can  make $m_{H_{u}}(m_Z)$ arbitrarily small, as it
happens in the MSSM.

\subsection{Fine--Tuning in  Simple Versions of `Mirage Mediation'}

\begin{figure}[!h]
\centerline{\CenterObject{\includegraphics[scale=0.9]{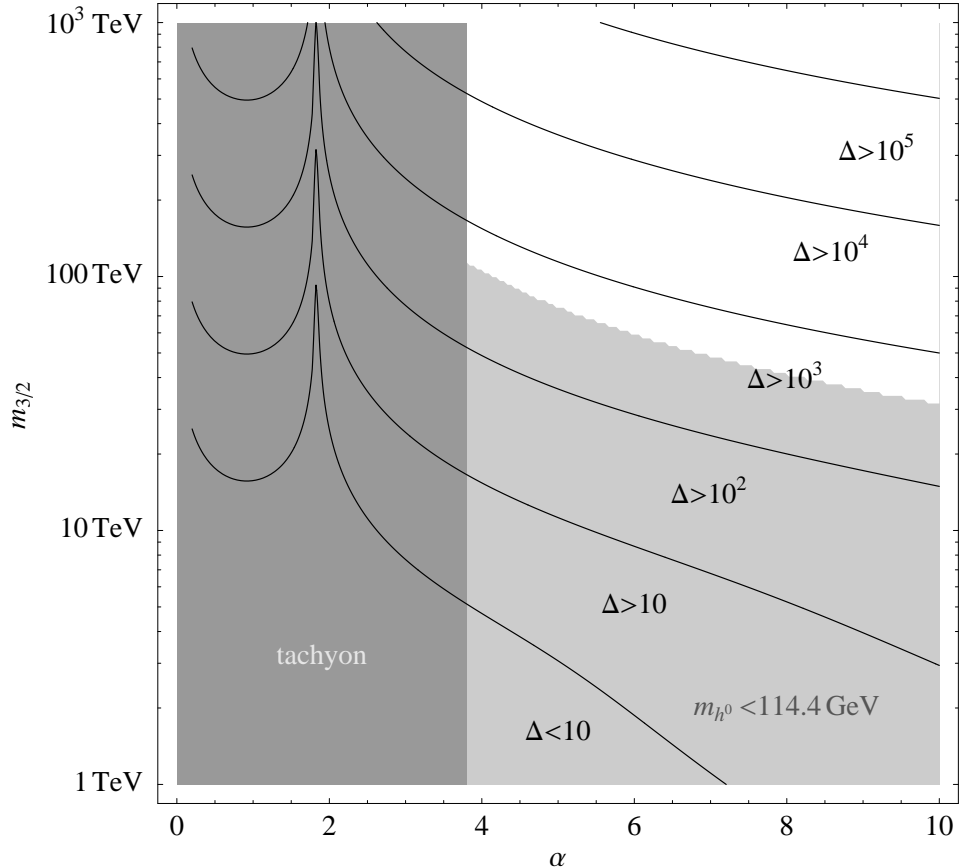}}}
\caption{
Fine--tuning $\Delta$ as a function of $\alpha$ and $m_{3/2}$ at $\tan\beta=5$
and $m_t=174\,\mathrm{GeV}$. 
The darker shaded area shows the presence of tachyons, while the lighter shaded
area is excluded by the LEP constraint on the Higgs mass. 
The electroweak symmetry breaking and the LEP  chargino mass
constraints   (cf.\ \cite{Falkowski:2005ck})
are not shown. }
\label{fig:Sensitivity}
\end{figure}

One may now ask what is the degree of fine--tuning typical to `mirage
mediation'. As a representative   example, let us  
consider `minimal mirage mediation' where the soft
mass parameters  are given by Eq.~\eqref{e.mmast}.
To quantify the fine--tuning, we will need a proper measure. A reasonable
measure of fine--tuning is given by a variation in the $Z$--mass upon a small
change in the SUSY GUT parameters. Specifically, we define the sensitivity
$\Delta_{\xi_i}$ of  $m_Z$ to the input parameters $\xi_i$ of the theory by
\cite{Barbieri:1987fn}:
\begin{equation}
 \Delta_{\xi_i}\,\frac{\delta\xi_i}{\xi_i} ~=~\frac{\delta m_{Z}^2}{m_Z^2}\;.
\end{equation}
In our setup, the relevant parameters are $\alpha$ and $m_{3/2}\equiv 16 \pi^2
M_s$. Then, the question is how much $m_Z$ changes if we perturb these
parameters while keeping $\tan\beta$,  $\mu$ and $B\mu$ fixed. From
Eq.~\eqref{eq:mZ2} we have 
\begin{equation}
 \Delta_\xi~=~\frac{2\xi}{m_{Z}^2}
 \frac{\D }{\D \xi}\frac{m_{H_d}^2-m_{H_u}^2\,\tan^2\beta}{\tan^2\beta-1}
\end{equation}
with $\xi$  being $\alpha$ or $m_{3/2}$, and $m_{H_{d,u}}^2$ evaluated at the 
electroweak scale.
For our purposes, it is convenient to define a mean sensitivity $\Delta$, 
\begin{equation}\label{eq:Sensitivity}
 \Delta~=~\sqrt{\Delta_\alpha^2+\Delta_{m_{3/2}}^2}\;.
\end{equation}
$\Delta$ is calculated by taking numerical derivatives of  $m_{H_u}^2$ and $m_{H_d}^2$
with respect to  $\alpha$ and $m_{3/2}$ at one loop. 
Our results are  presented in Fig.~\ref{fig:Sensitivity}.

These results can be understood as follows. Expressing the mass parameters in
Eq.~\eqref{eq:Kane} in terms of $\alpha$ and $M_s$  (Eq.~\eqref{e.mmast}), we
get
\begin{equation} 
\label{mzalpha}
 m_Z^2 ~\sim~ - 1.8\,\mu^2 + 4.5\, M_s^2 (\alpha^2 -3.7\, \alpha +3.1)  \;.
\end{equation}
Now, the degree of fine--tuning can be estimated analytically. 
We find that $\Delta_\alpha=0$ at $\alpha \simeq 2$ and $\Delta_{m_{3/2}}=0$ at $\alpha \simeq 2.4$
and 1.3. For the mean sensitivity,  we have
\begin{equation} 
\Delta \rightarrow {\rm min   ~~~at~~~} \alpha \simeq 2 \;,
\end{equation}
in which case $\Delta$ is $\leq 1$ for $\sim100\,\mathrm{GeV}$ soft masses. This
is also evident from Fig.~\ref{fig:Sensitivity}. On the other hand, at $\alpha
\simeq 2$ there is no electroweak symmetry breaking, i.e.\ Eq.~\eqref{mzalpha}
cannot be satisfied for any $\mu$ and $M_s$. Furthermore, the squarks and the
sleptons are tachyonic at the GUT scale so the boundary conditions are simply
ill--defined. An even more restrictive bound comes from the Higgs mass limit
which excludes large portions of the parameter space.

The main features of $\Delta (\alpha, m_{3/2})$ are quite transparent.
Fine--tuning increases rapidly with  $m_{3/2}$ since the gravitino mass sets the
scale of the soft masses. It also increases with $\alpha$ for two reasons.
First, larger $\alpha$ correspond to gravity dominated SUSY breaking and thus
larger gluino masses, and  second, they increase  the scale of the  soft masses.
The sharp decrease in the fine--tuning around $\alpha=2$ is a special property of
the `mirage mediation' soft terms. However, the area with $\Delta <100-1000$
(depending on $\tan\beta$) is excluded by the Higgs mass bound. 

It appears that although  `mirage mediation' has a nice qualitative feature
that the  gluino mass is suppressed, a combination of the EW symmetry breaking,
Higgs mass bound and absence of tachyons constraints typically requires a
relatively large degree of fine--tuning, similar to that of mSUGRA.

\subsection{Towards a Solution of the Fine--Tuning Problem}

In its simple incarnation, `mirage mediation' somewhat ameliorates
the MSSM fine--tuning problem, yet does not solve it.
There exist  regions in the parameter space where the fine--tuning is
small, but these are problematic for various reasons, in particular,
tachyonic boundary conditions at the GUT scale.

The above problems can perhaps  be circumvented in more general versions
of `mirage mediation'. In particular, given enough freedom in `effective modular 
weights', one can arrange for significant cancellations in  Eq.~\eqref{eq:Kane}
consistently with other constraints. This is a model--dependent issue and can be
studied only within particular   semirealistic      models realizing the MSSM on D--branes.

What is clear, however, is that `mirage mediation' has a robust feature that 
$M_3<M_2$   at the GUT scale and thus the fine--tuning is reduced
\cite{Roszkowski:1995cn,Kane:1998im,Bastero-Gil:1999gu}.
In particular, one can study the fine--tuning with respect to 
(to a large extent) model independent parameters such as the gaugino masses.
In `mirage mediation', the gaugino contribution to the $Z$--mass is given by
(cf.\ Eq.~\eqref{eq:Kane})
\begin{equation}
 \delta m_Z^2|_\mathrm{gaugino}
 ~\simeq~5.9\,M_3^2-0.4\,M_2^2
 ~\simeq~5.5\,M_s^2\,(\alpha-1.1)\,(\alpha-2.1)\;.
\end{equation} 
In the gravity mediation limit $\alpha\to\infty$, with $\alpha\,M_s$
fixed, the gaugino contribution to the $Z$--mass is at least of the order of
the soft masses $\alpha M_s$. On the other hand, in `mirage mediation'  this
contribution is reduced and can even be zero. 
That means that  a generic prediction of this scenario is that 
 the  usual  MSSM fine--tuning is reduced.

\section{Summary}

The scheme of `mirage mediation' has a number  of phenomenologically desirable
features. Most notably, the usual conflict between supergravity theories and
nucleosynthesis, known as the `gravitino/moduli problem', is resolved  since the
gravitino and moduli are  sufficiently heavy to decay before nucleosynthesis.

In this note, we have discussed the issue of fine--tuning associated with
obtaining the correct electroweak breaking scale in the `mirage  mediation'
scenario. We find that there exist regions in the parameter space where the
sensitivity  of $m_Z$  to the  input parameters $(\alpha,m_{3/2})$ is
considerably reduced. However, in  simple versions of `mirage mediation',
these regions are problematic since the corresponding GUT boundary  conditions
are tachyonic. This appears to be a model--dependent feature. Presumably, with a
more sophisticated choice of `effective modular weights' one can avoid tachyons
as well as satisfy the electroweak symmetry breaking conditions and the Higgs
mass bound.

An important model--independent feature of  `mirage mediation' is that it
predicts  $M_2>M_3$ at the high energy scale consistently with grand
unification. Therefore, the `mirage mediation' scheme ameliorates to some extent
the notorious MSSM fine--tuning.

\subsection*{Acknowledgments}

One of us (M.R.) would like to thank J\"orn Kersten for providing a
Mathematica interface to SOFTSUSY \cite{Allanach:2001kg,Skands:2003cj}, and
Ben Allanach, Sabine Kraml and Werner Porod for correspondence.
This work was partially supported by the European Union 6th Framework Program
MRTN-CT-2004-503369 `Quest for Unification' and MRTN-CT-2004-005104
`ForcesUniverse'.

\bibliography{Moduli}
\addcontentsline{toc}{section}{Bibliography}
\bibliographystyle{ArXiv}
\end{document}